\documentclass{llncs}

\usepackage{proof-dashed}
\usepackage{amsmath}
\usepackage{amssymb}
\usepackage{graphicx}
\usepackage{hyperref}
\usepackage{color}
\usepackage{listings}
\usepackage{stmaryrd}
\usepackage{comment}
\usepackage{todonotes}
\usepackage{import}

\newcommand{\m}[1]{\ensuremath{\mathsf{#1}}}
\newcommand{\pid}[1]{\m{#1}}
\newcommand{\pids}[1]{\til {\pid #1}}

\newcommand{\gentell}{\pid p\!: \pid q\, \code{<->}\, \pid r}

\newcommand{\emptyN}{\nil} 

\newcommand{\code}[1]{\texttt{#1}}
\newcommand{\nil}{\boldsymbol 0}

\newcommand{\com}[2]{#1\;\code{-\hspace{-0.3mm}>}\;#2}

\newcommand{\gencomf}{\com{\pid p.e}{\pid q.f}}
\newcommand{\sel}[3]{\com{#1}{#2 [#3]}}
\newcommand{\gensel}{\sel{\pid p}{\pid q}{l}}

\newcommand{\defs}{\mathcal{D}}
\newcommand{\pdefs}{\mathcal{B}}

\renewcommand{\merge}{\sqcup}
\newcommand{\cond}[3]{\m{if}\, #1 \, \m{then} \, #2 \, \m{else} \, #3}

\newcommand{\gencondE}{\cond{\pid p.e}{C_1}{C_2}}
\newcommand{\start}[2]{#1 \, \m{start} \, #2}
\newcommand{\astart}[2]{\m{start} \, #1 \triangleright #2}
\newcommand{\atells}[2]{{#1}!!{#2}}
\newcommand{\atellr}[2]{\arecv{#1}{#2}}

\newcommand{\genstart}{\start{\pid p}{\pid q}}

\newcommand{\gendef}{X(\wtil{\pid q}) = C}
\newcommand{\genpdef}{X(\pids q) = B}

\newcommand{\pn}{\m{pn}}

\newcommand{\callP}[2]{#1\langle #2 \rangle}

\newcommand{\gencallP}{\callP{X}{\pids p}}
\definecolor{light-gray}{gray}{0.928}

\newcommand{\pcont}{\mathtt{c}}
\newcommand{\til}{\tilde}
\newcommand{\wtil}{\widetilde}

\newcommand{\epp}[2]{[\![#1]\!]_{#2}}

\newcommand{\asend}[2]{{#1}!{#2}}
\newcommand{\arecv}[2]{#1?#2}
\newcommand{\actor}[3]{{#1} \triangleright_{#2} {#3}}
\newcommand{\parp}{\, \boldsymbol{|} \, }
\newcommand{\asel}[2]{{#1}\oplus#2}
\newcommand{\abranch}[2]{{#1}\&{#2}}

\newcommand{\acspar}{\hspace{0.8mm} | \hspace{0.8mm} }

\newcommand{\smallpar}[1]{\smallskip\noindent \textbf{\textit{#1.}}}

\makeatletter
\lstdefinestyle{mystyle}{
  basicstyle=\ttfamily%
             \lst@ifdisplaystyle\scriptsize\fi
}
\makeatother

\begin{document}

\title{Choreographies in Practice\thanks{Supported by \emph{CRC (Choreographies for Reliable and efficient
Communication software)}, grant no.\
DFF--4005-00304 from the Danish Council for Independent Research.}}
\author{Lu\'\i s Cruz-Filipe, Fabrizio Montesi}
\institute{
University of Southern Denmark, Department of Mathematics and Computer Science
\\
\email{$\{$lcf,fmontesi$\}$@imada.sdu.dk}
}

\maketitle

\begin{abstract}
Choreographic Programming is a development methodology for concurrent software that guarantees correctness by
construction. The key to this paradigm is to disallow mismatched I/O operations in programs, called
choreographies, and then mechanically synthesise distributed implementations in terms of standard process models via a 
mechanism known as EndPoint Projection (EPP).


Despite the promise of choreographic programming, there is still a lack of practical
evaluations that illustrate the applicability of choreographies to concrete computational problems with standard concurrent solutions.
In this work, we explore the potential of choreographies by using Procedural Choreographies (PC), a model that we 
recently proposed, to write distributed algorithms for sorting (Quicksort), solving linear equations (Gaussian 
elimination), and computing Fast Fourier Transform.
We discuss the lessons learned from this experiment, giving possible directions for the usage and future improvements
of choreography languages.
\end{abstract}

\section{Introduction}
\label{sec:intro}

Choreographic Programming is an emerging paradigm for the programming of concurrent software based on message 
passing~\cite{M13:phd}.
The key aspect of this paradigm is that programs are choreographies, i.e., global descriptions of communications based 
on an ``Alice and Bob'' notation inherited from security protocol notation~\cite{NS78}.
Since the syntax of such notation disallows the writing of mismatched I/O actions, choreographies always describe 
deadlock-free systems by construction.
Given a choreography, a distributed implementation can be projected automatically (synthesis) in terms of a process 
model. This transformation is typically called EndPoint Projection (EPP)~\cite{CHY12,CM13}.
If EPP is defined correctly, then the generated code behaves exactly as specified by the originating choreography, 
yielding a correctness-by-construction result: since a choreography cannot describe deadlocks, the generated 
process implementations must also be deadlock-free.
Previous works have presented formal models for capturing different aspects of choreographic programming, 
e.g., web services~\cite{CHY12,GGM15}, multiparty sessions and asynchrony~\cite{CM13}, runtime 
adaptation~\cite{DGGLM15}, modular development~\cite{MY13}, protocol compliance~\cite{CM13,CMS14}, and 
computational expressivity~\cite{ourstuff}.
More in general, looking also at other applications rather than just programming system implementations, choreographies 
have been investigated in the realms of type theory~\cite{HYC08}, automata theory~\cite{DY12,LTY15}, formal 
logics~\cite{CMSY15}, and service contracts~\cite{BZ07,QZCY07}.

Despite the rising interest in choreographies found in the communities of programming languages and concurrent 
computing, there is still a lack of evidence about what kind of nontrivial programs can actually be written with 
choreographic programming.
This is due to the young age of the paradigm~\cite{M15}. Indeed, most works on languages for choreographic programming still 
focus on showcasing representative toy examples (e.g.,~\cite{CHY12,CM13,chor:website,GGM15,M13:phd,MY13}), rather than 
giving a comprehensive practical evaluation of how known algorithms can be implemented using choreographies.
Here, we aim at contributing to filling this gap by investigating how choreographies can be used to 
tackle some common computational problems in the setting of concurrent programming.

Our investigation is based on the language of Procedural Choreographies (PC), and its corresponding process calculus of 
Procedural Processes (PP), which we presented in~\cite{ourPCstuff}. PC, introduced in~\S~\ref{sec:background}, is a 
choreography language developed with the aim of capturing divide-and-conquer concurrent algorithms, by extending 
previous choreography models with primitives for parameterised procedures.
Like several other choreography languages (e.g.,~\cite{CM13,MY13}), PC supports implicit parallelism by means of a
flexible semantics that allows non-interfering communications to take place in any order.

In this work, we provide an empirical evaluation of the expressivity of PC, by showing how
it can be used to program some representative and standard concurrent algorithms:
Quicksort (\S~\ref{sec:qs}), Gaussian elimination (\S~\ref{sec:gauss}), and Fast Fourier
Transform (\S~\ref{sec:fft}).
As a consequence of using choreographies, all these implementations are guaranteed to be deadlock-free.
We also illustrate how implicit parallelism has the surprising effect of automatically giving concurrent behaviour to traditional sequential implementations of these algorithms.
Our exploration brings us to the limits of the expressivity of PC, which arise when trying
to tackle distributed graph algorithms (\S~\ref{sec:graphs}), due to the lack of
primitives for accessing the structure of process networks, e.g., broadcasting a message
to neighbouring processes.


\section{Background}
\label{sec:background}

In this section, we recap the definition of Procedural Choreographies (PC), our
choreography language, and its main properties. We refer the reader to the original presentation of PC, 
in~\cite{ourPCstuff}, for a more comprehensive technical discussion.

\subsection{Procedural Choreographies}

The syntax of PC is introduced in Figure~\ref{fig:pc_syntax}.
A procedural choreography is a pair $\langle\defs,C\rangle$, where $C$ is a choreography
and $\defs$ is a set of procedure definitions.
\begin{figure}
\footnotesize
\begin{align*}
C ::={} & \eta;C \acspar I;C \acspar \nil
& \eta ::={} & \gencomf \acspar \gensel \acspar \genstart \acspar \gentell \\
\defs ::={} & \gendef, \defs \acspar \emptyset
& I ::={} & \gencondE \acspar \gencallP \acspar \nil
\end{align*}
\caption{Procedural Choreographies, Syntax.}
\label{fig:pc_syntax}
\end{figure}

Process names, ranged over by $\pid p,\pid q,\pid r,\ldots$, identify processes that
execute concurrently.
Each process $\pid p$ is equipped with a memory cell that stores a single value of a fixed type. In the remainder, we 
will omit such types in the language and our examples since they can always be inferred using the technique given 
in~\cite{ourPCstuff}.
Statements in a choreography can either be communication actions ($\eta$) or compound
instructions ($I$), and both can have continuations.
Term $\nil$ is the terminated choreography, which we sometimes omit.
The term $\nil;A$ is needed at runtime to capture the termination of procedure calls with
continuations.

Processes communicate via direct references (names) to each other. Communications are synchronous, as they are simple 
and suffice for our purposes here, but can be made asynchronous by adopting the asynchronous extension proposed 
in~\cite{CM13}.
In a value communication $\com{\pid p.e}{\pid q.f}$, process $\pid p$ sends the result of
evaluating expression $e$ to $\pid q$; the expression $e$ can contain the placeholder
$\pcont$, which is replaced at runtime with the data stored at process $\pid p$.
When $\pid q$ receives the value from $\pid p$, it applies to it the (total) function $f$,
of the form $\lambda x.e'$, replacing the formal parameter $x$ with the value sent by
$\pid p$.
The result of the computation will be stored in $\pid q$.
The expression $e'$, the body of $f$, can also contain the placeholder $\pcont$, allowing
it to read the contents of $\pid q$'s memory. We assume that expressions and functions are
written in a pure functional language, which we leave unspecified.

The selection term $\sel{\pid p}{\pid q}{l}$ is standard, as in session
types~\cite{HVK98}: $\pid p$ communicates to $\pid q$ its choice of label $l$.
Labels $l$ range over an infinite enumerable set.

In term $\genstart$, process $\pid p$ spawns the new process $\pid q$.
Process name $\pid q$ is bound in the continuation $C$ of $\genstart;C$. Also, after
executing $\genstart$, process $\pid p$ is assumed to be the only process who knows the name of process
$\pid q$. (Or, in other words, process $\pid p$ is the only process connected to $\pid q$.)
In order to propagate knowledge of $\pid q$ to other processes, PC includes the action
$\gentell$, read ``$\pid p$ introduces $\pid q$ and $\pid r$'', where $\pid p$, $\pid q$
and $\pid r$ are distinct.

In a conditional term $\gencondE$, process $\pid p$ evaluates expression $e$ to choose
between the possible continuations $C_1$ and $C_2$.

The set $\defs$ defines global procedures that can be invoked in choreographies.
Term $\gendef$ defines a procedure $X$ with body $C$, which can be used anywhere in
$\langle\defs,C\rangle$ -- in particular, inside the definitions of $X$ and other
procedures.
The names $\pids q$ are bound to $C$, and they are assumed to be exactly the free process
names in $C$.
The set $\defs$ contains at most one definition for each procedure name.
Term $\gencallP$ then invokes procedure $X$, instantiating its parameters with the
processes $\pids p$.
%

The semantics of PC, which we do not detail, is a reduction semantics that relies on two extra elements: 
a total state function that assigns to each process the value it stores, representing the local memory of processes; 
and, a connection graph that keeps track of which processes know (are connected to) which other 
processes~\cite{ourPCstuff}.
%
%
In particular, two processes can only communicate if there is an edge between them in the connection graph.
%
%
%
%
Therefore,
it is possible for choreographies to
deadlock (be unable to reduce) because of errors in the programming of communications: if two processes are supposed to 
communicate but they are not connected according to the connection graph, the choreography gets stuck.
This issue is addressed by a simple typing discipline that we do not discuss here (see~\cite{ourPCstuff}).
When a choreography is well-typed in PC, it is guaranteed to be deadlock-free.
\begin{theorem}[Deadlock-freedom~\cite{ourPCstuff}]
Let $\langle\defs,C\rangle$ be a well-typed procedural choreography. Then, $\langle\defs,C\rangle$ is deadlock-free.
\end{theorem}

\subsection{Procedural Processes}

Choreographies in PC are compiled into a distributed implementation represented in terms
of a process calculus: the calculus of Procedural Processes (PP).

The syntax of PP is reported in Figure~\ref{fig:pp_syntax}.
%
\begin{figure}[b]
\footnotesize
\begin{displaymath}
\begin{array}{rl}
B &{}::= \asend{\pid q}{e};B \acspar \arecv{\pid p}{f};B \acspar \atells{\pid q}{\pid r};B \acspar \atellr{\pid p}{\pid r};B \acspar \asel{\pid q}{l};B
\acspar \abranch{\pid p}{\{ l_i : B_i\}_{i\in I}};B \acspar \nil
\\[1ex]
& \acspar \astart{\pid q}{B_2};B_1 \acspar \cond{e}{B_1}{B_2};B \acspar \gencallP ;B \acspar \nil;B
\\[1ex]
N,M &{}::= \actor{\pid p}{v}{B} \quad | \quad N \parp M \quad | \quad \emptyN
\\[1ex]
\pdefs & {}::= \genpdef, \pdefs \acspar \emptyset
\end{array}
\end{displaymath}
\caption{Procedural Processes, Syntax.}
\label{fig:pp_syntax}
\end{figure}

A term $\actor{\pid p}{v}{B}$ is a process, where $\pid p$ is its name, $v$ is the value
stored in its memory cell, and $B$ is its behaviour.
Networks, ranged over by $N,M$, are parallel compositions of processes, where $\emptyN$ is
the inactive network.
Finally, $\langle\pdefs,N\rangle$ is a procedural network, where $\pdefs$ defines the
procedures that the processes in $N$ may invoke.
Values, expressions and functions are as in PC.

We comment on behaviours.
A send term $\asend{\pid q}{e};B$ sends the evaluation of expression $e$ to process $\pid q$, and then proceeds as $B$.
Term $\arecv{\pid p}{f};B$ is the dual receiving action: it receives a value from process
$\pid p$, combines it with the value in memory cell of the process executing the behaviour
as specified by $f$, and then proceeds as $B$.
Term $\atells{\pid q}{\pid r}$ sends process name $\pid r$ to $\pid q$ and process name
$\pid q$ to $\pid r$, making $\pid q$ and $\pid r$ ``aware'' of each other.
The dual action is $\atellr{\pid p}{\pid r}$, which receives a process name from $\pid p$
that replaces the bound variable $\pid r$ in the continuation.
Term $\asel{\pid q}{l};B$ sends the selection of a label $l$ to process $\pid q$.
Selections are received by the branching term $\abranch{\pid p}{\{ l_i : B_i\}_{i\in I}}$,
which can receive a selection for any of the labels $l_i$ and proceed according to $B_i$.
Branching terms must offer at least one branch.
Term $\astart{\pid q}{B_2};B_1$ starts a new process (with a fresh name) executing $B_2$,
proceeding in parallel as $B_1$.
The other terms are standard (conditionals, procedure calls, and termination), while
procedural definitions are stored globally as in PC.

Some terms bind names: $\astart{\pid q}{B_2};B_1$ binds $\pid q$ in $B_1$, and
$\atellr{\pid p}{\pid r};B$ binds $\pid r$ in $B$.
%
The semantics of PP is again a reduction semantics, capturing the intuitive description of
the operators given above.

\subsection{EndPoint Projection (EPP)}
\label{sec:ac_epp}
We now exhibit the compilation of procedural choreographies in PC to processes in PP.

\smallpar{Behaviour Projection}
We start by defining how to project the behaviour of a single process $\pid p$, a partial
function denoted $\epp{C}{\pid p}$.
The rules defining behaviour projection are given in Figure~\ref{fig:pc_epp}.
\begin{figure}[t]
\footnotesize
\begin{eqnarray*}
&&\epp{\gencomf;C}{\pid r} =
	\begin{cases}
		\asend{\pid q}{e};\epp{C}{\pid r} & \text{if } \pid r = \pid p \\
		\arecv{\pid p}{f};\epp{C}{\pid r} & \text{if } \pid r = \pid q \\
		\epp{C}{\pid r} & \text{else}
	\end{cases}
\quad
\epp{\gensel;C}{\pid r} =
	\begin{cases}
		\asel{\pid q}{l};\epp{C}{\pid r} & \text{if } \pid r = \pid p \\
		\abranch{\pid p}{\{ l : \epp{C}{\pid r} \}};\nil & \text{if } \pid r = \pid q \\
		\epp{C}{\pid r} & \text{else}
	\end{cases}
\\[1ex]
&&\epp{\gentell;C}{\pid s} = 
\begin{cases}
	\atells{\pid q}{\pid r};\epp{C}{\pid s} & \text{if } \pid s = \pid p
	\\
	\atellr{\pid p}{\pid r};\epp{C}{\pid s} & \text{if } \pid s = \pid q
	\\
	\atellr{\pid p}{\pid q};\epp{C}{\pid s} & \text{if } \pid s = \pid r
	\\
	\epp{C}{\pid s} & \text{else}
\end{cases}
\qquad
\begin{array}{l}
\epp{\gencallP;C}{\pid r} =
	\begin{cases}
		\callP{X_i}{\pids p};	\epp{C}{\pid r}
		& \text{if } \pid r = \pid p_i \\
		\epp{C}{\pid r} & \text{else}
	\end{cases}
\\
\epp{\nil}{\pid r} = \nil
\\
\epp{\nil;C}{\pid r} = \epp{C}{\pid r}
\end{array}
\\[1ex]
&&\epp{\gencondE;C}{\pid r} =
	\begin{cases}
		\cond{e}{\epp{C_1}{\pid r}}{\epp{C_2}{\pid r}}
		; \epp{C}{\pid r}
		& \text{if } \pid r = \pid p \\
		( \epp{C_1}{\pid r} \merge \epp{C_2}{\pid r} )
		;\epp{C}{\pid r} & \text{else}
	\end{cases}
\\[1ex]
&&\epp{\genstart;C}{\pid r} =
\begin{cases}
	\astart{\pid q}{\epp{C}{\pid q}};\epp{C}{\pid r}
	& \text{if } \pid r = \pid p
	\\
	\epp{C}{\pid r} & \text{else}
\end{cases}
\end{eqnarray*}
\caption{Procedural Choreographies, Behaviour Projection.}
\label{fig:pc_epp}
\end{figure}
Each choreography term is projected to the local action of the process that we are
projecting.
For example, for a communication term $\gencomf$, we project a send action if we are
projecting the sender process $\pid p$, a receive action if we are projecting the receiver
process $\pid q$, or we just proceed with the continuation otherwise.

The rule for projecting a conditional uses the standard (and partial) merging operator
$\merge$: $B\merge B'$ is isomorphic to $B$ and $B'$ up to branching, where the branches
of $B$ or $B'$ with distinct labels are also included~\cite{CHY12}.
Merging allows the process that decides a conditional to inform the other processes of its
choice later on, using selections~\cite{LGMZ08}.
Unlike in previous work, our conditionals have continuations, which have to be moved
inside the different branches if they do not coincide for all cases (see the example in
\S~\ref{sec:qs}).

Building on behaviour projection, we define how to project the set $\defs$ of procedure
definitions. Formally, the projection 
$\epp{\defs}{}$ is the component-wise extension of the projection of a single procedure, defined below.
\[\epp{\gendef}{} \ = \ 
X_1(\pids q)={\epp{C}{\pid q_1}},\ldots,
X_n(\pids q)={\epp{C}{\pid q_n}}
\qquad
\text{where }\pids q = \pid q_1,\ldots,\pid q_n
\,.
\]

\begin{definition}[EPP from PC to PP]
Given a procedural choreography $\langle\defs,C\rangle$ and a map of initial process values $\sigma$, the endpoint
projection $\epp{\defs,C,\sigma}{}$ is defined as the parallel composition of the
processes in $C$ with all global definitions derived from $\defs$:
\[
\textstyle\epp{\defs,C,\sigma}{}  \ = \ 
\left\langle\epp\defs{},\epp{C,\sigma}{}\right\rangle \ = \
\left\langle\epp\defs{},\prod_{\pid p \in \pn(C)} \actor{\pid p}{\sigma(\pid p)}{\epp{C}{\pid p}}\right\rangle
\]
where
$\epp{C,\sigma}{}$, the EPP of $C$ wrt state $\sigma$, is independent of $\defs$.
\end{definition}
Above, $\sigma$ is a total function mapping process names to their current values. For our purposes here, we will only 
consider a default mapping that assigns an initial undefined value to each process, and omit further discussions 
on $\sigma$ since it does not influence our presentation in any way (see~\cite{ourPCstuff} for details).

\smallpar{Properties}
EPP guarantees the typical operational correspondence between PC and PP: the projection of a choreography implements 
exactly the behaviour of the originating choreography. This implies, in
particular, that the projections of typable PC terms never deadlock.


\section{Quicksort}
\label{sec:qs}

In this section, we illustrate PC's capability of supporting the programming of divide-and-conquer algorithms, by providing a detailed implementation of (concurrent) Quicksort.

We begin by defining procedure \lstinline+split+, which splits the (non-empty) list stored at \lstinline+p+ among three processes: \lstinline+q<+, \lstinline+q=+ and \lstinline+q>+.
Before giving the code for \lstinline+split+, we describe the (standard) auxiliary functions and procedures that we are going to use.
We assume that all processes store objects of type \lstinline+List(T)+, where \lstinline+T+ is
some type. We also assume that these lists are implemented such that the following operations are supported and take constant time:  accessing the first element
(\lstinline+fst+); accessing the second element (\lstinline+snd+); checking that the length of a list is at most $1$ (\lstinline+short+); appending an element (\lstinline+add+); and, appending another list (\lstinline+append+).
This can be readily achieved, for example, by an implementation of linked lists with pointers
to the first, second and last node (and \lstinline+short+ simply checks where the pointer
to the second node is null).
We use the predicates \lstinline+fst<snd+ and \lstinline+fst>snd+ to test whether the first element
of the list at a process is, respectively, smaller or greater than the second element. Finally, the
procedure \lstinline+pop2+ (which we omit) removes the
second element from the list at its argument process.

We use the abbreviation \lstinline+p -> q1,...,qn[l]+ to signify that \lstinline+p+ sends
the label \lstinline+l+ to the processes \lstinline+q1+,\ldots,\lstinline+qn+, i.e., as an
abbreviation for the sequence of selections \lstinline+p -> q1[l]; ...; p -> qn[l]+.
We can now show the code for \lstinline+split+, reported in the following.




\begin{lstlisting}
split(p,q<,q=,q>) =
  if p.short then p -> q<,q=,q>[stop]; p.fst -> q=.add
  else if p.fst<snd then p -> q<[get]; p.snd -> q<.add; p -> q=,q>[skip]
       else if p.fst>snd then p -> q>[get]; p.snd -> q>.add; p -> q<,q=[skip]
            else p -> q=[get]; p.snd -> q=.add; p -> q<,q>[skip]
       ;
       pop2<p>; split<p,q<,q=,q>>
\end{lstlisting}

Procedure \lstinline+split+ starts by testing whether the list at process \lstinline+p+ is \lstinline+short+. If so, its element is stored in process \lstinline+q=+ and the procedure terminates.
Otherwise, we test whether the second element in the list is smaller, greater, or equal to the first element in the list, and add it to the respective process \lstinline+q<+, \lstinline+q=+, or \lstinline+q>+; then, we pop the second element of the list at \lstinline+p+ and recursively invoke \lstinline+split+.
When \lstinline+split+ terminates, we know that all elements in \lstinline+q<+ and \lstinline+q>+ are respectively smaller and greater than those in \lstinline+q=+.

Sending the label \lstinline+skip+ to the processes that will not receive messages in an iteration is required for projectability. (So that they know whether they will receive a value or not.)
Using \lstinline+split+ we can implement a robust version of Quicksort (in the sense that
it works with lists containing duplicates), the procedure \lstinline+QS+ below.
We use \lstinline+p start q1, ..., qn+ as a shortcut for the sequence \lstinline+p start q1; ...; p start qn+.
Observe that \lstinline+split+ is only called when \lstinline+p+ stores a non-empty list.

\begin{lstlisting}
QS(p) = if p.short then 0
        else p.start q<,q=,q>;
             split<p,q<,q=,q>>; QS<q<>; QS<q>>;
             q<.c -> p.id; q=.c -> p.append; q>.c -> p.append
\end{lstlisting}

\begin{figure}
  \centering
  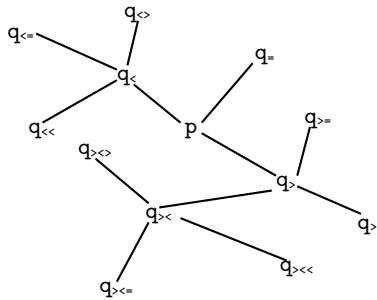
  \caption{Example of a network connection graph after some recursive calls of
    \lstinline+QS+.}
  \label{fig:luis-has-too-much-free-time}
\end{figure}

Procedure \lstinline+QS+ implements Quicksort using its standard recursive structure.
However, the recursive calls run completely in parallel here.
Indeed, the created processes \lstinline+q<+, \lstinline+q=+ and \lstinline+q>+ do not even have
references to each other, so they cannot exchange messages.
Therefore, the network of processes becomes tree-like, as exemplified in Figure~\ref{fig:luis-has-too-much-free-time}.

Applying EPP, we get the following process procedures, where we
simplified the projections of procedure definitions to include only the arguments that are actually used inside the procedures (see~\cite{ourPCstuff}).

\begin{lstlisting}
split_p(p,q<,q=,q>) =
  if short then q<\oplusstop; q=\oplusstop; q>\oplusstop; q=!fst
  else if fst<snd then q<\oplusget; q<!snd; q=\oplusskip; q>\oplusskip
       else if fst>snd then q>\oplusget; q>!snd; q<\oplusskip; q=\oplusskip
            else q=\oplusget; q=!snd; q<\oplusskip; q>\oplusskip
       ;
       pop2<p>; split_p<p,q<,q=,q>>

split_q<(p,q) = p&{stop: 0, get: p?add;split_q<(p,q), skip: split_q<(p,q)}

split_q=(p,q) = p&{stop: p?add, get: p?add;split_q=(p,q), skip: split_q=(p,q)}

split_q>(p,q) = p&{stop: 0, get: p?add;split_q>(p,q), skip: split_q>(p,q)}

QS_p(p) = if small then 0
          else (start q< |> split_q<<p,q<>; QS_p<q<>; p!c);
               (start q= |> split_q=<p,q=>; p!c);
               (start q> |> split_q><p,q>>; QS_p<q>>; p!c);
               q<?id; q=?append; q>?append
\end{lstlisting}

\begin{remark}
Our implementation of \lstinline+split+ is suitable in a context where communication is cheap, e.g., as in 
object-oriented programming and/or a multi-threaded application.
In architectures where communications are costly, it could be better to use a \lstinline+select+ function at
\lstinline+p+ to compute the lists of elements smaller than, equal to, or larger than the
pivot and send each of these in a single message to \lstinline+q<+, \lstinline+q=+ or
\lstinline+q>+, respectively.
\end{remark}

\section{Gauss Elimination}
\label{sec:gauss}

We now show how we can program the distributed resolution of systems of linear equations by Gaussian elimination.
Let $A\vec x=\vec b$ be a system of linear equations in matrix form; our procedure
\lstinline+gauss+ will transform this into an equivalent system $U\vec x=\vec y$, where
$U$ is an upper triangular matrix (so this system can be solved by direct substitution).
We use parameter processes $\pid a_{ij}$, with $1\leq i\leq n$ and $1\leq j\leq n+1$.
Each $\pid a_{ij}$ such that $1\leq i,j\leq n$ stores one value from the coefficient
matrix and $\pid a_{i,{n+1}}$ stores the independent term in one equation.
(Including the independent terms in the coefficient matrix substantially simplifies the notation, as Gaussian elimination treats the independent vector exactly as the columns
in the coefficient matrix.)
After execution, each $\pid a_{ij}$ stores the corresponding term in the new system.
For simplicity, we assume that the matrix $A$ is non-singular and numerically stable.

Implementing this algorithm in PC cannot be done directly, as our procedure \lstinline+gauss+ needs to take a variable number of parameters (the $\pid a_{ij}$).
However, it is straightforward to extend PC so that procedures can also take process
\emph{lists} as parameters, instead of only processes.
We succinctly describe how to do this.
\begin{description}
\item[Syntax of PC.] We extend parametric procedures with arguments that are lists of
  process names.
  In procedure calls, we can use standard list functions (e.g., head, tail) to manipulate
  these lists.
  These functions must be pure and take a list as their only argument.
  The processes in these lists are assumed to have all the same type, and process lists
  can only be used in procedure calls.
  In our examples, we will use uppercase letters to identify process lists and lowercase
  letters for normal process identifiers.
\item[Semantics of PC.] We assume that a procedure that is called with an empty list as
  one of its arguments is equivalent to the terminated process $\nil$.
\item[Connections.] We assume that the connections between processes are uniform wrt
  argument lists, i.e., if \lstinline+p+ is a process and \lstinline+A+ is a list of processes that
  are arguments to some procedure \lstinline+X+, then \lstinline+X+ requires/guarantees
  that \lstinline+p+ be connected to none or all of the processes in \lstinline+Q+.
  The type system presented in~\cite{ourPCstuff} can be trivially extended to check for
  this requirement.
\item[Syntax of PP.] We extend it in the same way as PC.
\item[Semantics of PP.] We assume that a procedure unfolds to $\nil$ if the process
  unfolding it does not occur inside its arguments.
\item[EndPoint Projection.] We project procedures as before, with one PP procedure for
  each argument of each PC procedure -- regardless of whether the argument is a single
  process or a list of processes.
  The merge operator $\merge$ also needs to be slightly adjusted; we explain how this is
  done in \S~\ref{sec:fft}, as this change is not required for the example in this
  section.
\end{description}

This extension preserves the properties of PC, PP, and the EPP from the former to the latter.

We use \lstinline+hd+ and \lstinline+tl+ to compute the head and tail of a list of
processes, respectively; \lstinline+fst+ and \lstinline+rest+, which take a
list of processes representing a matrix and return, respectively, the first row of the
matrix, or the matrix without its first row; and \lstinline+minor+, which removes both the
first row and the first column from a matrix.
(Formally, some of these functions would need the size of the rows as arguments, but we
omit these for simplicity.)
Each process uses standard arithmetic operations to combine its value with the one it receives.

The code of procedure \lstinline+gauss+ follows.
\begin{lstlisting}
gauss(A) = solve(fst(A));
           eliminate(fst(A),rest(A));
           gauss(minor(A))

solve(A) = divide_all(hd(A),tl(A)); set_to_one(hd(A))

divide_all(a,A) = divide(a,hd(A)); divide_all(a,tl(A))

divide(a,b) = a.c -> b.div

eliminate(A,B) = elim_row(A,fst(B)); eliminate(A,rest(B))

elim_row(A,B) = elim_all(tl(A),hd(B),tl(B)); set_to_zero(hd(B))

elim_all(A,m,B) = elim_one(hd(A),m,hd(B)); elim_all(tl(A),m,tl(B))

elim_one(a,m,b) = b start x; b: x <-> a; b: x <-> m;
                  a.c -> x.id; m.c -> x.mult; x.c -> b.minus

set_to_zero(a) = a start p; p.0 -> a.id
set_to_one(a) = a start p; p.1 -> a.id

\end{lstlisting}
Procedure \lstinline+solve+ divides the first equation by the pivot, obtaining the new
first equation in the reduced system.
Then, \lstinline+eliminate+ uses this row to perform an elimination step, setting the
first column of the coefficient matrix to zeroes.
The auxiliary procedure \lstinline+elim_row+ performs this step at the row level, using
\lstinline+elim_all+ to iterate through a single row and \lstinline+elim_one+ to perform
the actual computations.
The first row and the first column of the matrix are then removed in the recursive call,
as they will not change further.

This implementation follows the standard sequential algorithm for Gaussian elimination, as
described in, e.g., Algorithm~8.4 in~\cite{GGKK}.
However, the implicit parallelism in the semantics of choreographies allows it to run
concurrently.
We explain this behaviour by focusing on a concrete example.
Assume that $A$ is a $3\times 3$ matrix, so there are $12$ processes in total.
For legibility, we will write
\lstinline+b1+ for the independent term \lstinline+a14+ etc.;
\lstinline+A=\langlea11,a12,a13,b1,a21,a22,a23,b2,a31,a32,a33,b3\rangle+ for the matrix;
\lstinline+A1=\langlea11,a12,a13,b1\rangle+ for the first row
(likewise for \lstinline+A2+ and \lstinline+A3+);
and, \lstinline+A'2=\langlea22,a23,b2\rangle+
and likewise for \lstinline+A'3+.

Calling \lstinline+gauss(A)+ unfolds to
\begin{lstlisting}
solve(A1); eliminate(A1,\langleA2,A3\rangle);
solve(A'2); eliminate(A'2,A'3);
solve(\langlea33,b3\rangle)
\end{lstlisting}
or, unfolding \lstinline+eliminate+,
\begin{lstlisting}
solve(A1); elim_row(A1,A2); elim_row(A1,A3);
solve(A'2); elim_row(A'2,A'3);
solve(\langlea33,b3\rangle)
\end{lstlisting}

Unfolding \lstinline+solve(A1)+ is straightforward, leading to
\begin{lstlisting}
a11.c -> a12.div; a11.c -> a13.div; a11.c -> b1.div;
a11 start x1; x1.1 -> a11.id
\end{lstlisting}
and likewise for the remaining calls.
In turn, \lstinline+elim_row(A1,A2)+ becomes
\begin{lstlisting}
elim_all(\langlea12,a13,b1\rangle,a21,\langlea22,a23,b2\rangle); set_to_zero(a21)
\end{lstlisting}
which can be expanded to
\begin{lstlisting}
elim_one(a12,a21,a22); elim_one(a13,a21,a23); elim_one(b1,a21,b2);
set_to_zero(a21)
\end{lstlisting}
and we note that each of these procedure calls involves only communication between the
processes explicitly given as arguments.

Since all these procedures involve \lstinline+a21+, the semantics of choreographies
requires them to be executed in this order.
Likewise, the call to \lstinline+elim_row(A1,A3)+ must be executed afterwards (since it
also involves processes \lstinline+a11+ through \lstinline+a13+), and unfolds to a
sequential composition of procedure calls with \lstinline+a31+ as argument.

The interesting observation is that none of the processes intervening in
\lstinline+elim_row(A1,A3)+ occur in the expansion of \lstinline+solve(A'2)+.
In other words,
\begin{lstlisting}
elim_row(A1,A3); solve(A'2)
\end{lstlisting}
expands to
\begin{lstlisting}
elim_one(a12,a31,a32); elim_one(a13,a31,a33); elim_one(b1,a31,b3); set_to_zero(a31);
a21.c -> a22.div; a21.c -> a23.div; a21.c -> b2.div; a21 start x2; x2.1 -> a21.id
\end{lstlisting}
and the semantics of PC therefore allows the communications in the second line to be
interleaved with those in the first line in any possible way; in the terminology
of~\cite{ourstuff}, the calls to \lstinline+elim_row(A1,A3)+ and \lstinline+solve(A'2)+
run in parallel.

This corresponds to the implementation of Gaussian elimination with pipelined
communication and computation described in \S~8.3 of~\cite{GGKK}.
Indeed, as soon as any row has been reduced by all rows above it, it can apply
\lstinline+solve+ to itself and try to begin reducing the rows below it.
It is a bit surprising that we get such parallel behaviour by straightforwardly
implementing an imperative algorithm; the explanation is that the EndPoint Projection
encapsulates the part of determining which communications can take place in parallel, thus
removing this burden from the programmer.
In the next section, we will include a simple example of the EPP of a procedure with parameter lists.

\section{Fast Fourier Transform}
\label{sec:fft}

We now present a more complex example: computing the discrete Fourier transform of a
vector.
We refer the reader to \S~13.1 of~\cite{GGKK} for details.

\begin{definition}
  Let $\vec x=\langle x_0,\ldots,x_{n-1}\rangle$ be a vector of $n$ complex numbers.
  The \emph{discrete Fourier transform} of $\vec x$ is
  $\vec y=\langle y_0,\ldots,y_{n-1}\rangle$, where
  $y_j = \sum_{k=0}^{n-1}x_k\omega^{kj}$
  with $\omega=e^{2\pi i/n}$.
\end{definition}

Given $\vec x$, its discrete Fourier transform can be computed efficiently by the
\emph{Fast Fourier Transform} (FFT) as follows (Algorithm~13.1 in~\cite{GGKK}).
We assume $n$ to be a power of $2$; in the first call, $\omega$ has the value defined
earlier.

\begin{tabbing}
\textsf{procedure} R\_FFT($X$,$Y$,$n$,$\omega$) \\
\textsf{if} $n=1$ \= \textsf{then} $y_0=x_0$ \\
\> \textsf{else} \=
R\_FFT($\langle x_0,x_2,\ldots,x_{n-2}\rangle$,$\langle q_0,q_1,\ldots,q_{n/2}\rangle$,$n/2$,$\omega^2$) \+\+\\
R\_FFT($\langle x_1,x_3,\ldots,x_{n-1}\rangle$,$\langle t_0,t_1,\ldots,t_{n/2}\rangle$,$n/2$,$\omega^2$) \\
\textsf{for} $j=0$ \textsf{to} $n-1$ \textsf{do} \\
\quad $y_j=q_{(j \% \frac{n}{2})}+\omega^jt_{(j \% \frac{n}{2})}$
\end{tabbing}

To implement this procedure in PC, we need a way to communicate labels in
selections to a group of processes.
We do this by means of two auxiliary procedures \lstinline+gsel_then(p,Q)+ and
\lstinline+gsel_else(p,Q)+, where \lstinline+p+ is a process broadcasting a selection of
label \lstinline+then+ or \lstinline+else+, respectively, to all the processes in \lstinline+Q+.
In order for EPP to work correctly, we also need to extend the merge
operator $\merge$ slightly so as to recognize these procedures.
We give the definition of \lstinline+gsel_then+; this can trivially be adapted to any
other label.

\begin{lstlisting}
gsel_then(p,Q) = gsel1_then(p,hd(Q)); gsel_then(p,tl(Q))
gsel1_then(p,q) = p -> q[then]
\end{lstlisting}

The EPP of \lstinline+gsel_then+ looks as follows.

\begin{lstlisting}
gsel_then_p(p,Q) = gsel1_then_p(p,hd(Q)); gsel_then_p(p,tl(Q))
gsel_then_Q(p,Q) = gsel1_then_q(p,hd(Q)); gsel_then_Q(p,tl(Q))

gsel_then_p(p,q) = q\oplusl
gsel_then_q(p,q) = p&{then: 0}
\end{lstlisting}
The key aspect is that, during execution, the call to \lstinline+gsel_then_Q(p,Q)+ will
reduce to \emph{either} \lstinline+gsel1_then_q(p,hd(Q))+ \emph{or}
\lstinline+gsel_then_Q(p,tl(Q))+, as each process can only be on one of \lstinline+hd(Q)+ or
\lstinline+tl(Q)+!
So, while \lstinline+gsel_then_p(p,Q)+ essentially unfolds to a sequence of selections from
\lstinline+p+ to each of the processes in \lstinline+Q+, each local copy of
\lstinline+gsel_then_Q(p,Q)+ at a process \lstinline+q+$\in$\lstinline+Q+ unfolds exactly to
the reception of one selection from \lstinline+p+.

We will use (without specifying them) the following auxiliary procedures.
\begin{itemize}
\item \lstinline+intro(n,m,P)+, where \lstinline+n+ introduces \lstinline+m+ to every
  process in \lstinline+P+ (defined similarly to \lstinline+gsel_then+ above)
\item \lstinline+power(n,m,nm)+, where at the end \lstinline+nm+ stores the result of
  exponentiating the value in \lstinline+m+ to the power of the value stored in
  \lstinline+n+ (see~\cite{ourstuff} for a possible implementation in a sublanguage of
  PC).
\end{itemize}

Before we present our implementation of FFT, we point out the one major difference wrt the
algorithm R\_FFT reported above: we are not able to create a variable number of fresh processes and pass
them as arguments to other procedures (corresponding to the auxiliary vectors $\vec q$ and
$\vec t$).
Therefore, we have to use the result vector $\vec y$ to store the result of the recursive
calls, and then create two auxiliary processes inside each iteration of the final
\textsf{for} loop.

\begin{lstlisting}
fft(X,Y,n,w) = if n.is_one
               then gsel_then(n,join(X,Y)); n -> w[then]; base(hd(X),hd(Y))
               else gsel_else(n,join(X,Y)); n -> w[else];
                    n start n'; n.half -> n'; intro(n,n',Y);
                    w start w'; w.square -> w'; intro(w,w',Y);
                    n: n' <-> w; w: n' <-> w';
                    fft(even(X),half1(Y),n',w');
                    fft(odd(X),half2(Y),n',w');
                    n' start wn; n': w <-> wn; power(n',w,wn);
                    w start wj; w.1 -> wj; intro(w,wj,Y);
                    combine(half1(Y),half2(Y),wn,w,wi)

base(x,y) = x.c -> y

combine(Y1,Y2,wn,w,wj) = combine1(hd(Y1),hd(Y2),wn,wj);
                         w.c -> wj.mult;
                         combine(tl(Y1),tl(Y2),wn,w,wj)

combine1(y1,y2,wn,wj) = y1 start q; y1.c -> q; y1: q <-> y2;
                        y2 start t; y2.c -> t; y2: t <-> y1; y2: t <-> wj;
                        q.c -> y1; wj.c -> t.mult; t.c -> y1.add;
                        q.c -> y2; wn.c -> t.mult; t.c -> y2.add

\end{lstlisting}

The level of parallelism in this implementation is suboptimal, as the two recursive calls
to \lstinline+fft+ both use \lstinline+n'+ and \lstinline+w'+; by duplicating these
processes, however, these calls are able to run in parallel exactly as in the previous
example.
(We chose the current formulation for simplicity.)
Process \lstinline+n'+ is actually the main orchestrator of the whole execution.

\section{Graphs}
\label{sec:graphs}

Another prototypical application of distributed algorithms is graph problems.
In this section, we focus on a simple example (broadcasting a token to all nodes of a
graph) and discuss the limitations of implementing these algorithms in PC.

The idea of broadcasting a token in a graph is very simple: each node receiving the token
for the first time should communicate it to all its neighbours.
The catch is that, in PC, there are no primitives for accessing the connection graph
structure from within the language.
Nevertheless, we can implement our simple example of token broadcasting if we assume that
the graph structure is statically encoded in the set of available functions over parameters of
procedures.
To be precise, assume that we have a function \lstinline+neighb(p,V)+, returning the
neighbours of \lstinline+p+ in the set of vertices \lstinline+V+.
(The actual graph is encapsulated in this function.)
We also use \lstinline-++- and \lstinline+\+ for appending two lists and computing the set
difference of two lists.
We can then write a procedure \lstinline+broadcast(P,V)+, propagating a token from every
element of \lstinline+P+ to every element of \lstinline+V+, as follows.
\begin{lstlisting}
broadcast(P,V) = bcast(hd(P),neighb(hd(P),V));
                 broadcast(tl(P)++neighb(hd(P),V),V\neighb(hd(P),V))

bcast(p,V) = bcast_one(p,hd(V)); bcast(p,tl(V))

bcast_one(p,v) = p.c -> v.id
\end{lstlisting}
Calling \lstinline+broadcast(\langlep\rangle,G)+, where \lstinline+G+ is the full set of
vertices of the graph and \lstinline+p+ is one vertex, will broadcast \lstinline+p+'s
contents to all the vertices in the connected component of \lstinline+G+ containing
\lstinline+p+.
Furthermore, implicit parallelism again ensures that each node will start broadcasting as
soon as it receives the token, independently of the remaining ones.

This approach is however not very satisfactory as a graph algorithm, since it requires
encoding the whole graph in the definition of \lstinline+broadcast+; furthermore, it does
not generalise easily to more sophisticated graph algorithms.
Adding primitives for accessing the network structure at runtime is not simple, as it would influence the definitions 
of EPP and the type system of PC~\cite{ourPCstuff} (which we omitted in this presentation).
We leave this as an interesting direction for future work, which we plan to pursue in order to be able to implement 
more sophisticated graph algorithms, e.g., for computing a minimum spanning tree.
%

\section{Related Work}
\label{sec:related}

To the best of our knowledge, this is the first experience report on using
choreographic programming for writing 
real-world, complex computational algorithms.

The work nearest to ours is the evaluation of the Chor language~\cite{M13:phd},
an implementation of the choreographic 
programming model in~\cite{CM13}. Chor supports multiparty sessions (similar to channels in the 
$\pi$-calculus~\cite{MPW92}) and their mobility, which recalls introductions in PC. Chor is evaluated by encoding some 
representative examples from Service-Oriented Computing, such as distributed authentication and streaming, but none of 
the presented examples cover interesting algorithms as in here.

Previous works based on Multiparty Session Types (MPST)~\cite{HYC08} have explored
the use of choreographies as protocol specifications for the coordination of message exchanges in some real-world 
scenarios~\cite{DY11,NY15,YDBH10}. Differently from our approach, these works fall back to a standard process calculus 
model for defining implementations. Instead, our programs are choreographies.
As a consequence, programming the composition of separate algorithms in PC is done on the level of choreographies, 
whereas in MPST composition requires using the low-level process calculus.
Also, our choreography model is arguably 
much simpler and more approachable by newcomers, since much of the expressive power of PC comes from allowing 
parameterised procedures, a standard feature of most programming languages. The key twist in PC is that parameters are 
process names.

\section{Conclusions}
\label{sec:conclusions}

We have reported our experience with the writing of some representative concurrent
algorithms in the paradigm of choreographic programming.

What have we learned from this experience?

First, that choreographies make it easy to produce a simple concurrent implementation of a sequential algorithm. 
This is obtained by choosing process identifers to maximise the effect of implicit parallelism.
Then, EPP takes care of generating the concrete separate programs and the required I/O
actions to implement the described behaviour.
This is a striking difference from how concurrent algorithms usually differ from their
sequential counterparts.
Although we do not necessarily get the most efficient possible distributed algorithm, this
automatic concurrency is a pleasant property to observe.

The second interesting realisation is that it is relatively easy to implement nontrivial algorithms in choreographies. 
We exemplified this point with our implementations of Gaussian elimination and Fast Fourier Transform.
This is an important deviation from the typical use of toy examples, of limited practical
significance, that characterises previous works in this programming paradigm.


In conclusion, we showed that the current state of choreographic programming can already be used for addressing complex 
real-world computational problems. We also identified a future direction for extending the paradigm towards settings 
that require accessing the structure of process networks, such as some algorithms on graphs.

\bibliographystyle{abbrv}
\bibliography{biblio}

\begin{thebibliography}{10}

\bibitem{BZ07}
M.~Bravetti and G.~Zavattaro.
\newblock Towards a unifying theory for choreography conformance and contract
  compliance.
\newblock In M.~Lumpe and W.~Vanderperren, editors, {\em Proc.\ of SC}, volume
  4829 of {\em LNCS}, pages 34--50. Springer, 2007.

\bibitem{CHY12}
M.~Carbone, K.~Honda, and N.~Yoshida.
\newblock Structured communication-centered programming for web services.
\newblock {\em ACM Trans.\ Program.\ Lang.\ Syst.}, 34(2):8, 2012.

\bibitem{CM13}
M.~Carbone and F.~Montesi.
\newblock Deadlock-freedom-by-design: multiparty asynchronous global
  programming.
\newblock In R.~Giacobazzi and R.~Cousot, editors, {\em Proc.\ of POPL}, pages
  263--274. {ACM}, 2013.

\bibitem{CMS14}
M.~Carbone, F.~Montesi, and C.~Sch{\"{u}}rmann.
\newblock Choreographies, logically.
\newblock In P.~Baldan and D.~Gorla, editors, {\em Proc.\ of {CONCUR}}, volume
  8704 of {\em LNCS}, pages 47--62. Springer, 2014.

\bibitem{CMSY15}
M.~Carbone, F.~Montesi, C.~Sch{\"{u}}rmann, and N.~Yoshida.
\newblock Multiparty session types as coherence proofs.
\newblock In L.~Aceto and D.~de~Frutos{-}Escrig, editors, {\em Proc.\ of
  CONCUR}, volume~42 of {\em LIPIcs}, pages 412--426. Schloss Dagstuhl, 2015.

\bibitem{chor:website}
Chor.
\newblock {Programming Language}.
\newblock \url{http://www.chor-lang.org/}.

\bibitem{ourstuff}
L.~Cruz-Filipe and F.~Montesi.
\newblock Choreographies, computationally.
\newblock {\em CoRR}, abs/1510.03271, 2015.
\newblock Submitted for publication.

\bibitem{ourPCstuff}
L.~Cruz-Filipe and F.~Montesi.
\newblock Choreographies, divided and conquered.
\newblock {\em CoRR}, abs/1602.03729, 2016.
\newblock Submitted for publication.

\bibitem{DY11}
P.-M. Deni{\'e}lou and N.~Yoshida.
\newblock Dynamic multirole session types.
\newblock In T.~Ball and M.~Sagiv, editors, {\em Proc.\ of POPL}, pages
  435--446. {ACM}, 2011.

\bibitem{DY12}
P.-M. Deni{\'e}lou and N.~Yoshida.
\newblock Multiparty session types meet communicating automata.
\newblock In {\em Proc.\ of ESOP}, LNCS, pages 194--213. Springer-Verlag, 2012.

\bibitem{GGM15}
M.~Gabbrielli, S.~Giallorenzo, and F.~Montesi.
\newblock Applied choreographies.
\newblock {\em CoRR}, abs/1510.03637, 2015.

\bibitem{GGKK}
A.~Grama, A.~Gupta, G.~Karypis, and V.~Kumar.
\newblock {\em Introduction to Parallel Computing}.
\newblock Pearson, 2003.
\newblock 2nd edition.

\bibitem{HVK98}
K.~Honda, V.~Vasconcelos, and M.~Kubo.
\newblock Language primitives and type disciplines for structured
  communication-based programming.
\newblock In C.~Hankin, editor, {\em Proc.\ of ESOP}, volume 1381 of {\em
  LNCS}, pages 22--138. Springer, 1998.

\bibitem{HYC08}
K.~Honda, N.~Yoshida, and M.~Carbone.
\newblock Multiparty asynchronous session types.
\newblock In G.~C. Necula and P.~Wadler, editors, {\em Proc.\ of POPL}, pages
  273--284. {ACM}, 2008.

\bibitem{LGMZ08}
I.~Lanese, C.~Guidi, F.~Montesi, and G.~Zavattaro.
\newblock Bridging the gap between interaction- and process-oriented
  choreographies.
\newblock In {\em Proc. of SEFM}, pages 323--332. {IEEE}, 2008.

\bibitem{LTY15}
J.~Lange, E.~Tuosto, and N.~Yoshida.
\newblock From communicating machines to graphical choreographies.
\newblock In S.~K. Rajamani and D.~Walker, editors, {\em Proc.\ of POPL}, pages
  221--232. {ACM}, 2015.

\bibitem{MPW92}
R.~Milner, J.~Parrow, and D.~Walker.
\newblock A calculus of mobile processes, {I and II}.
\newblock {\em Information and Computation}, 100(1):1--40,41--77, Sept. 1992.

\bibitem{M13:phd}
F.~Montesi.
\newblock {\em Choreographic Programming}.
\newblock Ph.{D}. thesis, IT University of Copenhagen, 2013.
\newblock \url{http://fabriziomontesi.com/files/m13\_phdthesis.pdf}.

\bibitem{M15}
F.~Montesi.
\newblock Kickstarting choreographic programming.
\newblock {\em CoRR}, abs/1502.02519, 2015.

\bibitem{MY13}
F.~Montesi and N.~Yoshida.
\newblock Compositional choreographies.
\newblock In P.~R. D'Argenio and H.~C. Melgratti, editors, {\em Proc.\ of
  CONCUR}, volume 8052 of {\em LNCS}, pages 425--439. Springer, 2013.

\bibitem{NS78}
R.~M. Needham and M.~D. Schroeder.
\newblock Using encryption for authentication in large networks of computers.
\newblock {\em Commun.\ ACM}, 21(12):993--999, Dec. 1978.

\bibitem{NY15}
N.~Ng and N.~Yoshida.
\newblock Pabble: parameterised scribble.
\newblock {\em Service Oriented Computing and Applications}, 9(3-4):269--284,
  2015.

\bibitem{DGGLM15}
M.~D. Preda, M.~Gabbrielli, S.~Giallorenzo, I.~Lanese, and J.~Mauro.
\newblock Dynamic choreographies -- safe runtime updates of distributed
  applications.
\newblock In T.~Holvoet and M.~Viroli, editors, {\em Proc.\ of COORDINATION},
  volume 9037 of {\em LNCS}, pages 67--82. Springer, 2015.

\bibitem{QZCY07}
Z.~Qiu, X.~Zhao, C.~Cai, and H.~Yang.
\newblock Towards the theoretical foundation of choreography.
\newblock In C.~L. Williamson, M.~E. Zurko, P.~F. Patel{-}Schneider, and P.~J.
  Shenoy, editors, {\em Proc.\ of WWW}, pages 973--982. {ACM}, 2007.

\bibitem{YDBH10}
N.~Yoshida, P.~Deni{\'{e}}lou, A.~Bejleri, and R.~Hu.
\newblock Parameterised multiparty session types.
\newblock In C.~L. Ong, editor, {\em Proc.\ of FOSSACS}, volume 6014 of {\em
  LNCS}, pages 128--145. Springer, 2010.

\end{thebibliography}

\end{document}